\documentclass{WileyMSP-template}
\usepackage[numbers,sort&compress,comma,square]{natbib}
\usepackage{rotating}
\usepackage{booktabs}
\usepackage{pdflscape}
\usepackage{float}
\usepackage{graphicx}
\usepackage{booktabs}
\usepackage{tabularx}
\usepackage{array}

\newcolumntype{C}{>{\centering\arraybackslash}X}

\begin{document}

\pagestyle{fancy}
\rhead{\includegraphics[width=2.5cm]{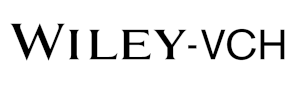}}

\title{Fabrication-Aware Design of a Hybrid Metasurface-Bragg Mirror for Low-Noise Precision Optics}

\maketitle

% Author: Please give full first and last names for authors and include * after the name of all corresponding authors

\author{Mika Gaedtke$^{\dagger,*}$}
\author{Christian Kranhold$^{\dagger}$}
\author{Markus Walther}
\author{Falk Eilenberger}
\author{Stefanie Kroker}
\author{Thomas Siefke}

% Dedication

\dedication{}

% Affiliations: Please provide adacemic titles (Prof. or Dr.) for all authors where applicable, and include an institutional email address for all corresponding authors
% Affiliations: Please provide academic titles (Prof. or Dr.) for all
% authors where applicable and include an institutional email address
% for all corresponding authors.
\begin{affiliations}

Mika Gaedtke, Prof. Dr. Stefanie Kroker \\
Institut für Halbleitertechnik, Technische Universität Braunschweig,
Hans-Sommer-Str. 66, 38106 Braunschweig, Germany \\
Laboratory for Emerging Nanometrology,
Langer Kamp 6a-b, 38106 Braunschweig, Germany \\

Christian Kranhold, Markus Walther, Dr. Falk Eilenberger,
Dr. Thomas Siefke \\
Friedrich Schiller University, Institute of Applied Physics,
Albert-Einstein-Str. 15, 07745 Jena, Germany \\

Dr. Falk Eilenberger, Dr. Thomas Siefke \\
Fraunhofer Institute for Applied Optics and Precision Engineering IOF,
Albert-Einstein-Str. 7, 07745 Jena, Germany \\

Prof. Dr. Stefanie Kroker \\
Physikalisch-Technische Bundesanstalt,
Bundesallee 100, 38116 Braunschweig, Germany \\

Dr. Thomas Siefke \\
Ernst Abbe University of Applied Sciences Jena,
Carl-Zeiss-Promenade 2, 07745 Jena, Germany \\

$^{*}$ Corresponding author:
mika.gaedtke@tu-braunschweig.de (M.G.)

$^{\dagger}$ These authors contributed equally to this work.

\end{affiliations}

% Keywords: Please provide a minimum of three and a maximum of seven keywords, separated by commas


% Abstract should be written in the present tense and impersonal style (i.e., avoid we), and be at most 200 words long
\begin{abstract}

   High-reflectivity coatings for precision interferometry must simultaneously minimize optical losses and thermally driven displacement noise. Dielectric Bragg mirrors provide robust high reflectance but rely on thick multilayer coatings, whereas metasurface mirrors provide high reflectance and low noise but are sensitive to fabrication-induced deviations. A fabrication-aware photonic design framework is introduced that combines a resonant single-layer metasurface, an etch-stop layer, an antiresonant spacer, and a reduced Bragg reflector. Fabrication effects, including geometric tolerances and line-edge roughness, are explicitly considered in the optical design. Full-wave electromagnetic simulations are combined with a truncated-Gaussian Monte Carlo analysis to determine the performance distribution under the assumed fabrication conditions. The ideal metasurface design achieves a modeled non-reflected power of $1-R<10^{-5}$. After roughness-aware reoptimization, the modeled non-reflected power remains below $1.04\cdot10^{-4}$ at $80\%$ fabrication yield. For the cryogenic ETpathfinder gravitational-wave testbed, two Bragg layer pairs reduce the non-reflected power of the complete stack to approximately $6.6\,\mathrm{ppm}$. The estimated thermal displacement-noise amplitude spectral density is $1.97\cdot10^{-20}\,\mathrm{m/\sqrt{Hz}}$ at $100\,\mathrm{Hz}$. The architecture connects fabrication robustness, optical performance, and thermal-noise reduction within a single design framework.

\end{abstract}

% Text: Please use section headings and subheadings as specified below. For communications, all section headings apart from Experimental Section should be removed
% Please make the first reference to a display item bold: \textbf{Figure 1}
% Do not abbreviate Figure, Equation, etc.; display items are always singular, i.e., Figure 1 and 2.
% Equations are always singular, i.e., Equation 1 and 2, and should be inserted using the {equation} environment, not as graphics
% Please do not use footnotes in the text, additional information can be added to the Reference list.

\section{Introduction}

High-reflectivity mirrors are essential components of precision interferometric systems, including ultrastable laser cavities, optical frequency metrology, and gravitational-wave detectors. In these systems, thermally driven fluctuations of the mirror substrates and coatings lead to displacement noise limiting the achievable sensitivity or frequency stability \cite{Numata:04,Kessler:12,PhysRevD.111.062002,granata2020a}. Mirror technologies for precision metrology must therefore combine extremely high reflectance and low optical absorption with minimal thermally induced noise. 

Conventional high-reflectivity dielectric mirrors consist of alternating layers of materials with high and low refractive indices. An increase in the number of layers increases the reflectance of such a mirror but also increases the thermal noise originating from the coating \cite{harry2002}. A high refractive-index contrast can reduce the required number of layers, but the achievable noise amplitude is fundamentally linked to the mechanical loss of the coating materials \cite{Levin1998}. The development of low-loss amorphous and crystalline coating materials has demonstrated that reducing mechanical dissipation can substantially improve the performance of precision interferometers, although optical absorption, scalability, and fabrication complexity remain important constraints \cite{granata2020a, Cole:13, cole2023a}. 

Resonant dielectric metasurfaces offer an alternative approach to high-reflectivity Bragg mirrors. Near-unity reflectance can be achieved by a periodic arrangement of subwavelength, high-index structures within a single structured layer \cite{kuznetsov2024a, Brueckner:10}. These structures generally exhibit a very low thermal noise amplitude compared to conventional coatings \cite{kroker2017a, dickmann2018c} and have been implemented in high-finesse optical cavities \cite{dickmann2023a}. However, the resonant optical response is highly sensitive to the geometrical parameters of the nanostructure. Deviations introduced during lithography, etching, and layer deposition can therefore considerably reduce the reflectance obtained. In addition to variations of nominal dimensions, line-edge roughness can modify the effective refractive-index profile and further degrade the optical response \cite{siefke2018a, siefke2024a}. Statistical tolerance analysis and uncertainty quantification are consequently required to distinguish nominal performance from the performance expected after fabrication \cite{Schmitt:19}.

Hybrid metasurface–multilayer mirrors have recently been investigated as alternatives to conventional high-reflectivity coatings. Hartmann et al. assess geometrical tolerance ranges through parameter sweeps \cite{Hartmann2026HybridMirror}, while Pandey et al. develop a Brownian-noise calculation framework and apply it to an architecture with the metasurface embedded beneath a Bragg reflector \cite{Pandey2026}. Here, we explicitly connect statistical fabrication performance to the design of the supporting Bragg coating. Line-edge roughness is incorporated through an effective smoothing model, and simultaneous geometrical and material variations are evaluated using a Monte Carlo analysis. Roughness-aware reoptimization and the resulting performance distribution determine the selection of the supporting coating at a specified modeled fabrication yield. The optical response is independently verified using rigorous coupled-wave analysis and finite-element simulations, and the Brownian, thermo-elastic, and thermo-refractive noise contributions of the complete stack are evaluated. The cryogenic ETpathfinder interferometer \cite{Utina_2022} provides a demanding application example for this fabrication-aware photonic design framework.

\section{Hybrid mirror concept}

\begin{figure}[h!]
    \centering
    \begin{minipage}[c]{0.49\textwidth}
    \caption{
        Schematic image of the hybrid metasurface--Bragg mirror concept. The integrated etalon consists of (1) a resonant periodic metasurface, (2) an etch-stop layer, (3) an anti-resonant spacer, (4) a Bragg mirror and (5) a substrate. The metasurface provides a dominant reflectance contribution, while the Bragg mirror compensates the residual transmission remaining under realistic fabrication conditions. 
    }
    \label{fig:hybrid_concept}
    \end{minipage}
    \hfill
    \begin{minipage}[c]{0.49\textwidth}
        \centering
        \includegraphics[width=\linewidth]{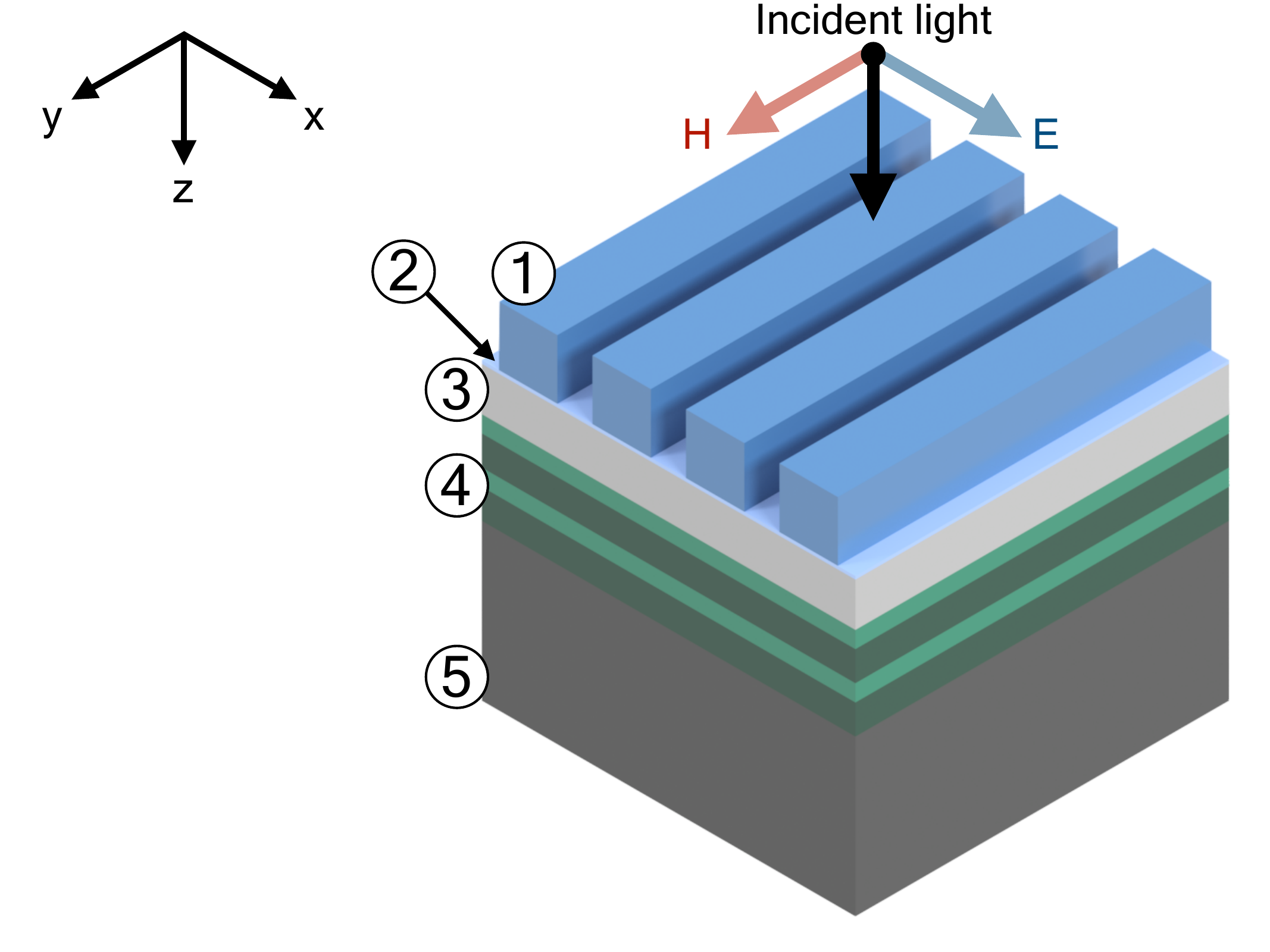}
    \end{minipage}
\end{figure}

\textbf{Figure \ref{fig:hybrid_concept}} illustrates the proposed hybrid mirror architecture. The mirror forms an integrated thin-film structure consisting of a resonant dielectric metasurface, a thin etch-stop layer, an antiresonant spacer, a reduced Bragg reflector, and a supporting substrate. The resonant metasurface provides the dominant reflectance contribution. The function of the underlying Bragg mirror is to suppress the residual power transmitted through the structured layer, particularly when fabrication-induced deviations reduce its nominal reflectance. 
The spacer between the metasurface and Bragg coating fulfills two related functions. Its optical thickness is selected to establish an antiresonant phase condition for the integrated structure, while its geometrical thickness limits near-field interaction between the resonant nanostructure and the rear coating. The thin etch-stop layer prevents overetching into the spacer during fabrication. In contrast to configurations in which two reflectors form a macroscopic cavity \cite{dickmann2023a}, all components of the present architecture are incorporated into a compact coating stack. The following sections first establish the optical response and fabrication tolerance of the metasurface and subsequently determine the number of Bragg coating layers required for the complete hybrid mirror.

\section{Design}

    \subsection{Optical design of the metasurface}
\begin{figure*}[h!]
            \centering
            \includegraphics[width=0.8\linewidth]{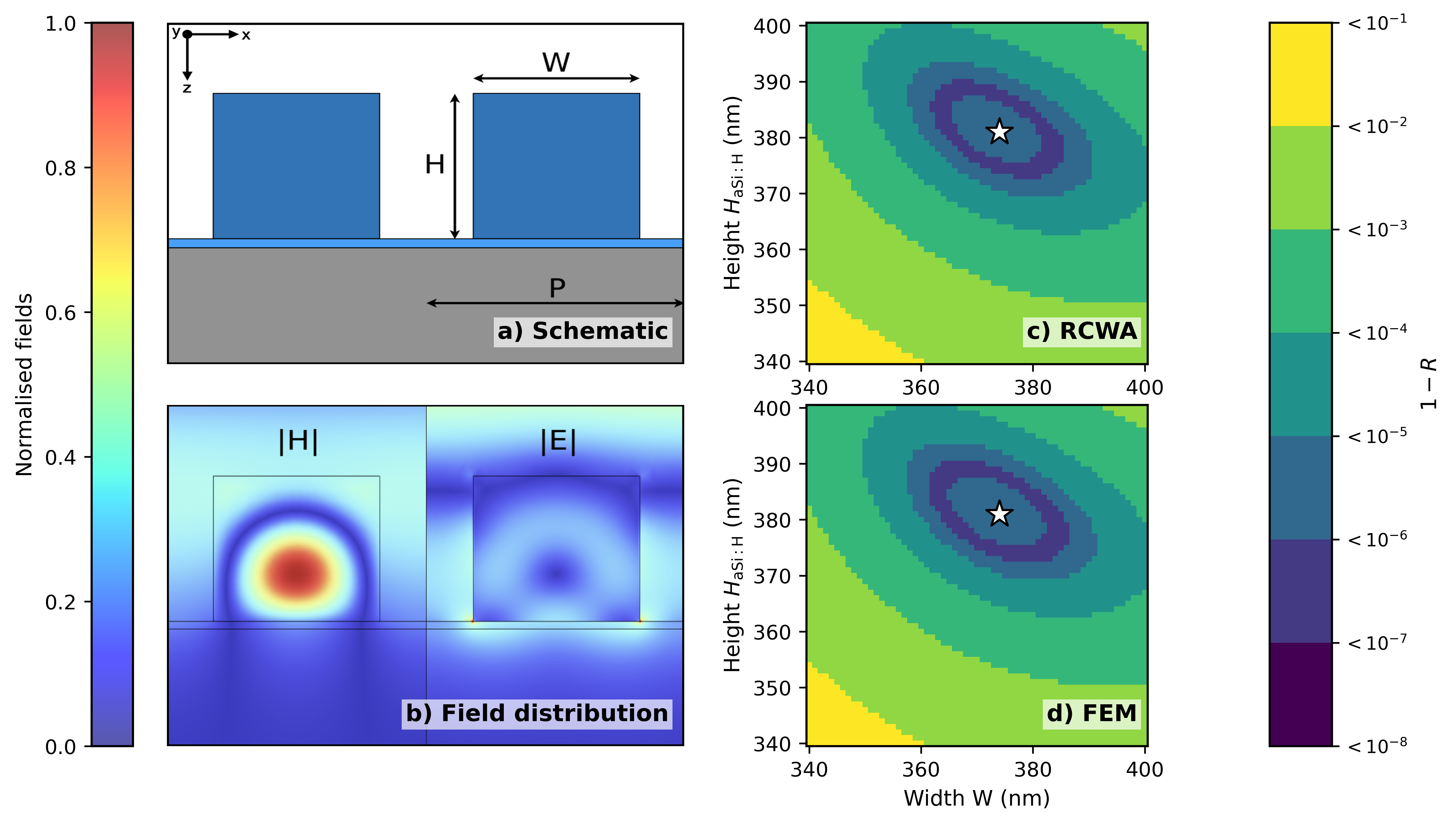}
            \caption{Design of the periodic metasurface. The design wavelength is $\mathrm{\lambda_0 = 1550\,nm}$. The period of the unit cell is $\mathrm{599\,nm}$. The metasurface consists of 1D-periodic ridges made from aSi:H on top of an etch-stop layer made from $\mathrm{Al_2O_3}$ on top of a $\mathrm{SiO_2}$ spacer. \textbf{a)} Schematic of the metasurface. W is the width, H the height and P the period of the metasurface. \textbf{b)} Normalized magnetic and electric fields inside the metasurface. \textbf{c)} RCWA simulation. The star marks the selected configuration with $\mathrm{W = 374\,nm}$ and $\mathrm{H = 381\,nm}$. \textbf{d)} FEM simulation. The results confirm the selected geometry determined through RCWA.}
            \label{fig:reflection_map}
        \end{figure*}

 The front reflector of the proposed hybrid mirror is made up of a periodic array of hydrogenated amorphous silicon (aSi:H) ridges. Hydrogenated amorphous silicon was chosen over the non-hydrogenated variant because of its substantially lower optical absorption at the design wavelength of $\mathrm{\lambda_0 = 1550\,nm}$. All material parameters used for simulations and calculations can be found in \textbf{Table \ref{table:matparam}}. Extinction coefficients achievable with hydrogenated amorphous silicon are up to two orders of magnitude lower than for conventional amorphous silicon \cite{birney2018amorphous,franta2013advanced, Takei:14}. Low absorption of the metasurface is of significant importance for the performance of the hybrid mirror. Residual transmission through the metasurface can be compensated by the Bragg mirror, whereas the power absorbed within the metasurface is irreversibly lost. Non-hydrogenated amorphous silicon as the metasurface material would result in absorption values of around $\mathrm{40\,ppm}$ for the proposed design. The design of the metasurface is shown in \textbf{Figure \ref{fig:reflection_map}}, a schematic of the metasurface geometry is shown in Figure \ref{fig:reflection_map} a). The simulated stack consists of the aSi:H layer on top of a $\mathrm{20\,nm}$ thick amorphous $\mathrm{Al_2O_3}$ etch-stop layer and a semi-infinite $\mathrm{SiO_2}$ substrate. In the final design, the latter becomes the finite spacer that separates the metasurface from the Bragg mirror. Each unit cell of the metasurface is described by the period $P$, the width of the ridges $W$ and their height $H$. The structure is illuminated by plane waves at normal incidence, with the magnetic field parallel to the ridges (see also Figure \ref{fig:hybrid_concept}). The normalized magnetic- and electric-field amplitudes of the selected design are shown in Figure \ref{fig:reflection_map} b), showing a strongly confined magnetic mode inside of the metasurface ridges.

        The geometrical parameters were optimized using rigorous coupled-wave analysis (RCWA) \citep{moharam1981a, jin2020a} in combination with gradient-free optimization \citep{nevergrad}. Results for the final configuration are shown in Figure \ref{fig:reflection_map} c). Width and height of the final structure were fixed at $\mathrm{W = 374\,nm}$ and $\mathrm{H=381\,nm}$, indicated by the star in Figure \ref{fig:reflection_map} c). An independent FEM simulation was performed to verify the geometry, the results are shown in Figure \ref{fig:reflection_map} d). The predicted reflectances of both simulations show very good agreement and are $\mathrm{R_{RCWA} \approx 0.9999962}$ and $\mathrm{R_{FEM} \approx 0.9999953}$, respectively. This corresponds to a difference of $\mathrm{0.9\,ppm}$. Across the common high-reflectance region, defined by $\mathrm{1-R \leq 10^{-5}}$, the mean absolute difference in reflectance is $\mathrm{0.63\,ppm}$. Small residual differences are consistent with the distinct numerical discretizations employed by the two methods: FEM approximates the electromagnetic fields on a finite spatial mesh, whereas RCWA represents the fields and permittivity using a finite number of Fourier harmonics \cite{Solano:14}.

    \subsection{Impact of line-edge roughness on metasurface performance}

        \begin{figure*}[h!]
            \centering
            \includegraphics[width=0.95\linewidth]{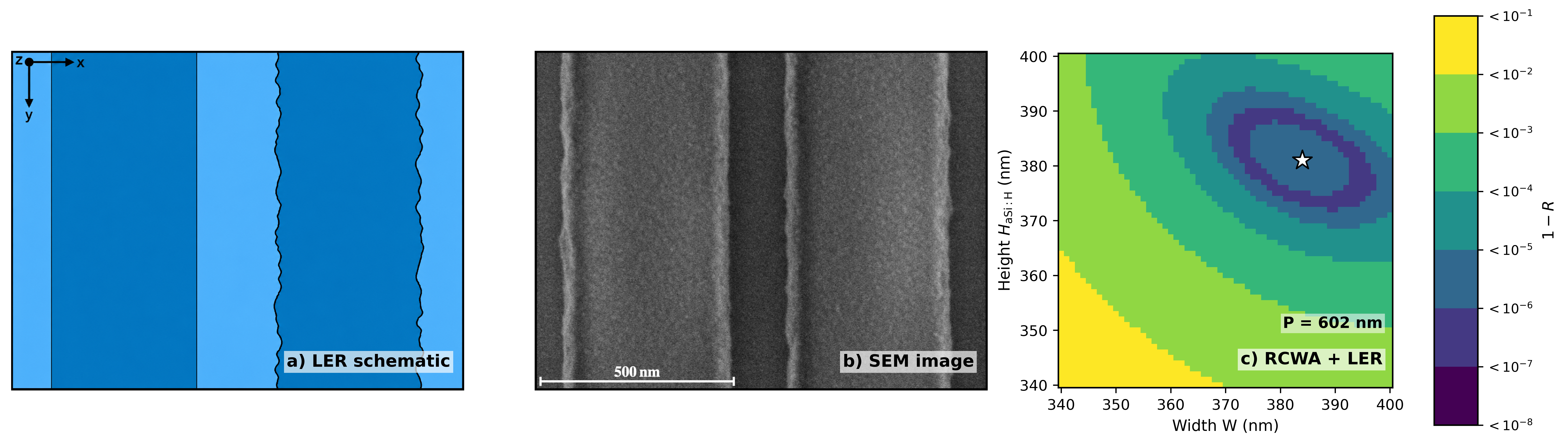}
            \caption{Reoptimized metasurface design accounting for line-edge roughness (LER). \textbf{a)} Schematic comparison between ideal and rough ridge boundaries. The roughness is exaggerated for better visibility. \textbf{b)} Representative SEM image showing LER in a layer of electron-beam-sensitive resist on silicon. \textbf{c)} RCWA simulation of the metasurface with LER. The roughness was implemented as a Gaussian smoothing of the refractive index profile with a full width at half maximum of $\mathrm{6.8\,nm}$. The reoptimized design has a period of $\mathrm{P = 602\,nm}$, a ridge width of $\mathrm{W = 384\,nm}$ and a ridge height of $\mathrm{H = 381\,nm}$.  }
            \label{fig:reflection_map_ler}
        \end{figure*}
        
        The results presented in the previous section assume a perfect structure, including perfectly sharp and straight material interfaces. In practice, lithography and subsequent pattern transfer steps will inherently lead to line-edge roughness (LER) \cite{10.1117/12.848236}. This corresponds to a stochastic variation of the ridge boundaries along the ridge direction. A schematic comparison between ideal and rough ridge boundaries is shown in \textbf{Figure \ref{fig:reflection_map_ler}} a). Figure \ref{fig:reflection_map_ler} b) shows a representative SEM image of LER in an electron-beam sensitive resist layer on top of silicon. In order to consider this effect in the simulations, a simplified model was used that describes LER as sigmoidal smoothing of the refractive index profile at the metasurface boundaries \cite{siefke2018a}. The LER is modeled by applying a Gaussian smoothing of the refractive index profile with a full width at half maximum of $\mathrm{6.8\,nm}$, which was estimated based on the standard deviations of prior lithography and etching processes \cite{ siefke2018a, siefke2024a}. Resolving the randomly varying ridge boundaries would require large three-dimensional simulations and a statistical evaluation of multiple roughness realizations. The effective model predominantly captures the influence of high-spatial-frequency roughness components and is used as an upper boundary estimate of the influence of LER. The metasurface geometry presented in the previous section was reoptimized with the LER model included in the RCWA simulations, the resulting reflectance map is shown in Figure \ref{fig:reflection_map_ler} c). The selected reoptimized design has a period of $\mathrm{P = 602\,nm}$, a ridge width of $\mathrm{W = 384\,nm}$ and a ridge height of $\mathrm{H = 381\,nm}$.

        In the following section, both the original design and LER optimized design are subjected to a statistical fabrication-tolerance analysis to compare their expected fabrication yields.
        
    \subsection{Fabrication tolerance analysis of the metasurface}

    Electron-beam lithography, layer deposition and the subsequent pattern transfer introduce deviations from the intended geometrical parameters. While in the previous section, LER was treated as a systematic smoothing of the refractive-index profile, the remaining fabrication tolerances are considered as statistical variations. A Monte Carlo analysis \cite{Metropolis01091949} was performed to evaluate their influence on the optical performance of the metasurface. Input parameters are repeatedly sampled randomly within uncertainty intervals estimated from underlying fabrication processes and the resulting geometries are evaluated. This allows simultaneous variations and interactions between multiple parameters to be considered \cite{Schmitt:19}. 
    
    Our analysis includes variations of the period, ridge width, ridge height, thickness of the etch-stop layer and side-wall angle for a periodic unit cell of the structure. Additionally, variations of the real part of the refractive index of all materials are also considered. All deviations were sampled from a zero-centered Gaussian distribution truncated at $\mathrm{\pm 2}$ standard deviations. These assumed standard deviations are $\mathrm{0.25\,nm}$ for the period, $\mathrm{12.5\,nm}$ for the ridge width, $\mathrm{5\,nm}$ for the ridge height, $\mathrm{2.5\,nm}$ for the etch-stop layer thickness and $\mathrm{2.5\,}$° for the side-wall angle. The relative standard deviation for the refractive indices is $\mathrm{0.15\,\%}$.
    
    The original design and LER-optimized design were subjected to the same statistical analysis, where the same randomly generated parameter variations were applied to both designs. In total, 20000 realizations were evaluated for each design. Furthermore, the same LER model, which corresponds to a Gaussian smoothing with a full width at half maximum of $\mathrm{6.8\,nm}$, was included in both cases. This allows the influence of the LER reoptimization to be compared under identical fabrication conditions.

        \begin{figure*}[h!]
            \centering
            \includegraphics[width=1\linewidth]{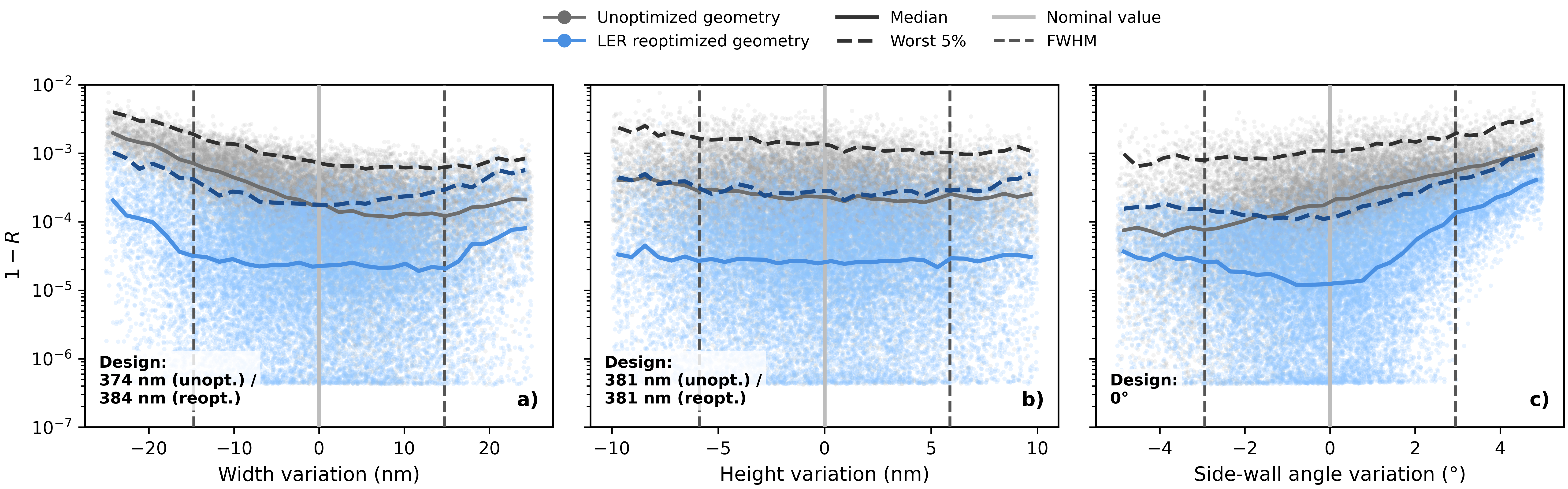}
            \caption{Monte Carlo tolerance analysis of the metasurface including fabrication uncertainties. The unoptimized design is compared to the LER reoptimized design. The total non-reflected power is shown in dependence of the \textbf{a)} ridge-width variation, \textbf{b)} height variation of the ridges and underlying etch-stop layer and \textbf{c)} side-wall angle variation. Gray and blue points represent the 20000 geometries for the original and LER reoptimized design, respectively. Solid colored lines show the medians, while dashed colored lines show the corresponding worst $\mathrm{5\,\%}$ of the realizations. Dashed vertical lines indicate the full width at half maximum of the corresponding input distributions.}
            \label{fig:mc_3panel}
        \end{figure*}

    \textbf{Figure \ref{fig:mc_3panel}} shows the results of the fabrication-tolerance analysis. The total non-reflected power, including residual transmission and absorption, is shown in dependence of variations of the metasurface geometry. Each point shown in Figure \ref{fig:mc_3panel} represents one Monte Carlo realization. Although the results are plotted against one fabrication parameter at a time, all remaining geometrical and material parameters were varied simultaneously. The results show a significant improvement of the non-reflected power for all geometrical variations for the LER reoptimized design. For very small non-reflected powers, all geometries are limited by absorption loss.

        \begin{figure}[h!]
            \centering
            \begin{minipage}[c]{0.49\textwidth}
            \caption{
                Cumulative distributions of the total non-reflected power obtained from the Monte Carlo fabrication-tolerance analysis. The original geometry is compared to the LER reoptimized geometry. With a $80\,\%$ probability the unoptimized design reaches $\mathrm{1-R = 6.22\cdot 10^{-4}}$ while the LER optimized design reaches $\mathrm{1-R = 1.04\cdot 10^{-4}}$. 
            }
            \label{fig:mc_summary}
            \end{minipage}
            \hfill
            \begin{minipage}[c]{0.49\textwidth}
                \centering
                \includegraphics[width=\linewidth]{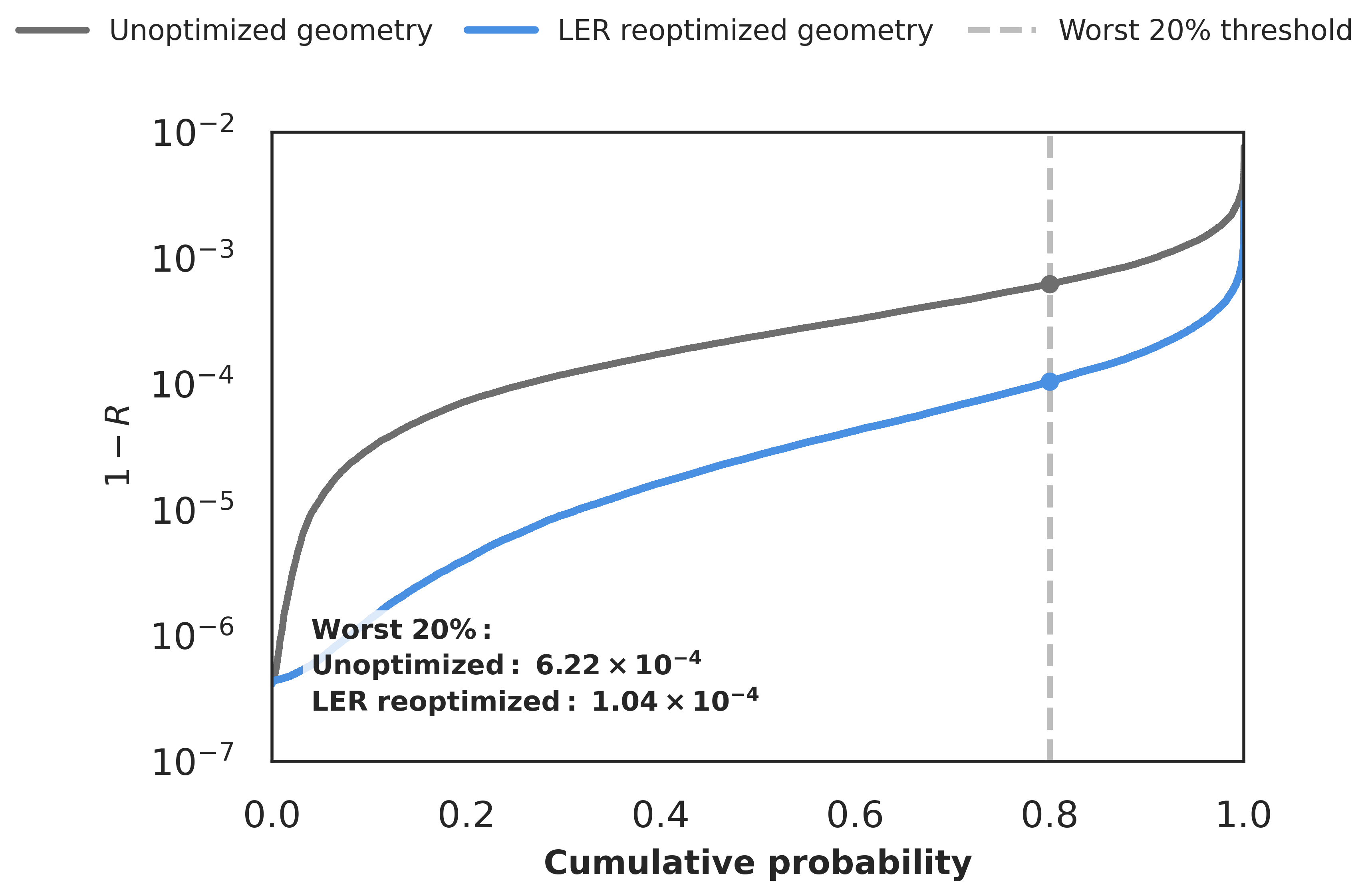}
            \end{minipage}
        \end{figure}        
            
    The complete distributions are compared in \textbf{Figure \ref{fig:mc_summary}}. The curves show the fraction of simulated geometries for which the non-reflected power remains below a given threshold value. At a cumulative probability of $\mathrm{80\,\%}$, the original geometry reaches $\mathrm{1-R = 6.22\cdot 10^{-4}}$, while for the LER reoptimized geometry this value is reduced to $\mathrm{1-R = 1.04\cdot 10^{-4}}$. The reoptimization therefore reduces the optical loss for this threshold by a factor of approximately 6. These results demonstrate that the reoptimization considering LER improves the nominal optical performance as well as the expected fabrication yield. The non-reflected power of $\mathrm{1-R = 1.04\cdot 10^{-4}}$ at a yield of $\mathrm{80\,\%}$ will be used as the expected performance of the metasurface for the final design. This threshold of $\mathrm{80\,\%}$ was chosen arbitrarily. Choosing a lower fabrication yield probability would increase the expected reflectance of the fabricated devices, while in turn lowering the odds of actually achieving the desired reflectance. In the following section, suitable choices for the Bragg coating and spacer thickness are discussed.

\subsection{Selection of the supporting Bragg coating and spacer}

The selection of a Bragg-coating material system is application specific and does not represent a universally optimal solution. In particular, the relative performance of the investigated coatings depends on the operating temperature and wavelength, the optical requirements, and the assumed cryogenic material parameters. The present comparison is therefore performed for the ETpathfinder case study introduced in Section~\ref{sec:pathfinder}. Among the configurations satisfying $\mathrm{T_{res} \leq 10^{-5}}$, $\mathrm{aSi/HfO_2}$ yields the lowest estimated coating Brownian noise and was selected as the supporting reflector. Details of the comparison are provided in Appendix~B. The final number of layer pairs is subsequently determined from the residual-transmission requirement of the complete hybrid mirror.

The $\mathrm{SiO_2}$ spacer thickness was selected to satisfy the antiresonance condition while suppressing direct near-field interaction between the metasurface and the first Bragg layer. Reflection phases of both reflectors were numerically evaluated and used for the selection of the spacer thickness. Further details on the calculation are provided in Appendix \ref{Appendix:spacer}. The smallest possible thickness, for which a further increase no longer meaningfully changes the simulated residual transmission, is $\mathrm{d_{spacer}\approx 279\,nm}$.

\section{Performance Evaluation under ETpathfinder Operating Conditions}
\label{sec:pathfinder}

ETpathfinder is a cryogenic research and development facility established to develop technologies for future gravitational-wave observatories, with particular relevance to the Einstein Telescope. The facility implements complete interferometer configurations on an approximately $\mathrm{10\,m}$ scale, allowing cryogenic test masses, optical components, suspension and cooling systems, and advanced noise-reduction techniques to be investigated under system-level low-noise conditions \cite{Utina_2022}. Its initial design considers two complementary operating configurations: one using silicon test masses at $\mathrm{18\,K}$ and a wavelength of $\mathrm{1550\,nm}$, and another operating at $\mathrm{123\,K}$ and $\mathrm{2090\,nm}$. In the present work, the $\mathrm{1550\,nm}$ and $\mathrm{18\,K}$ configuration serves as a demanding application example for evaluating the proposed hybrid mirror architecture.
For this case study, the relevant optical and mechanical parameters are a residual-transmission target of $\mathrm{T_{res}\leq10^{-5}}$, a Gaussian beam radius of $\mathrm{2.2\,mm}$, and a target displacement sensitivity below

$\mathrm{\approx2\cdot10^{-19}\,m/\sqrt{Hz}}$ at a mechanical frequency of $\mathrm{100\,Hz}$. These system-level requirements determine the number of supporting Bragg layer pairs and therefore translate the general hybrid-mirror concept into an application-specific design.

\begin{figure*} [h!]
            \centering
            \includegraphics[width=0.9\linewidth]{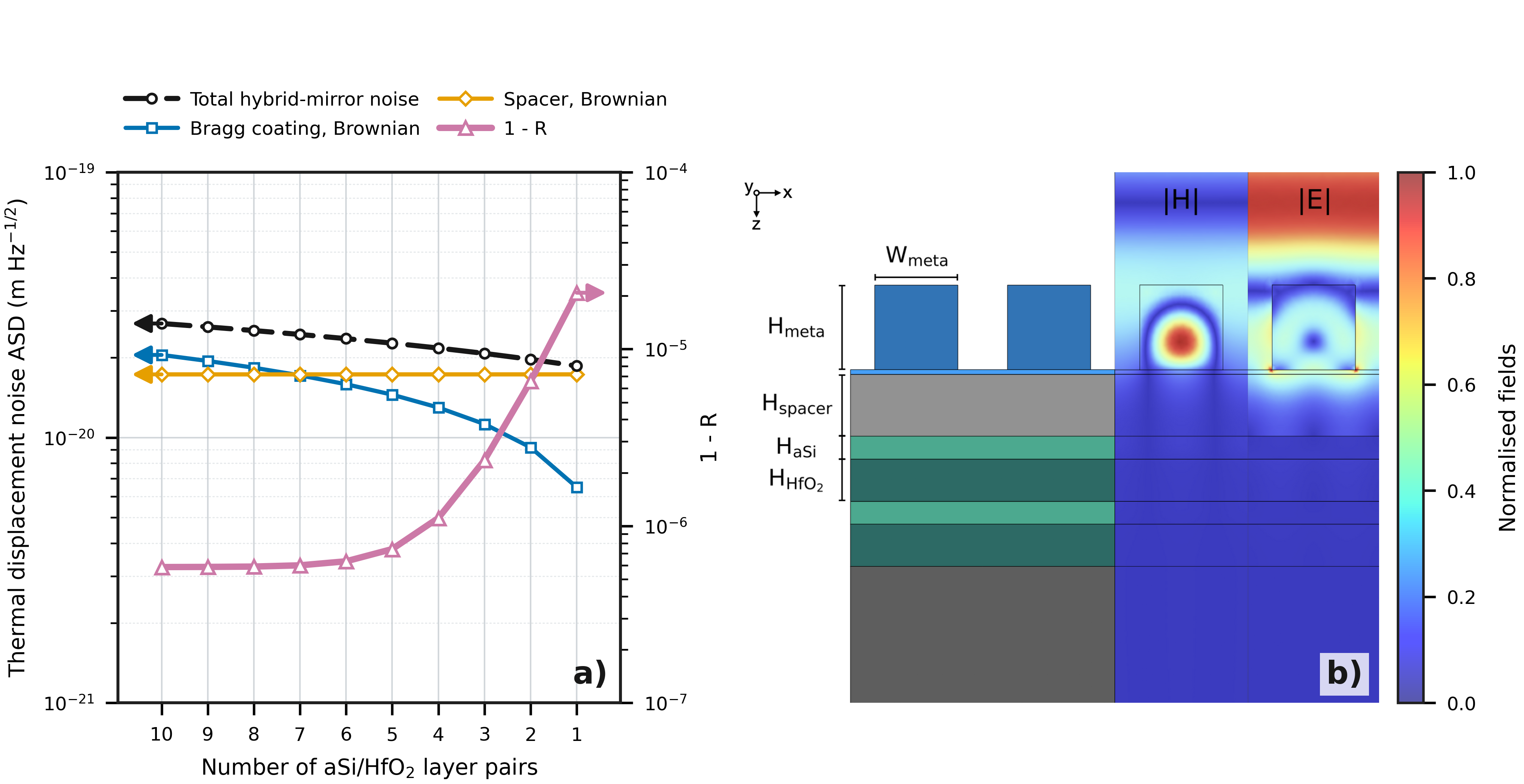}
            \caption{
            Influence of the number of $\mathrm{aSi/HfO_2}$ Bragg layer pairs on the optical and thermal-noise performance of the hybrid mirror at a mechanical frequency of $\mathrm{100\,Hz}$. \textbf{a)} Total thermal displacement noise and Brownian noise contributions of the Bragg coating and spacer at a mechanical frequency of 100 Hz, together with the non-reflected power. The arrows indicate the corresponding vertical axes. Two layer pairs represent the smallest Bragg stack for which the non-reflected power of 6.6 ppm remains below the ETpathfinder requirement of 10 ppm. For 7 or more layer pairs, the non-reflected power is entirely limited by the absorption of the structure. \textbf{b)} Schematic of the selected hybrid mirror with two Bragg layer pairs, together with the normalized magnetic- and electric-field amplitudes throughout the complete structure. The final geometrical parameters are: $\mathrm{P=602\,nm}$, $\mathrm{W_{meta} = 384\,nm}$, $\mathrm{H_{meta} = 381\,nm}$, $\mathrm{H_{spacer} = 279\,nm}$, $\mathrm{H_{aSi} = 102.67\,nm}$, $\mathrm{H_{HFO_2} = 188.84\,nm}$.}
                \label{fig:noise_DBR_sweep}
        \end{figure*}

The final mirror design consists of the LER-optimized metasurface, a $\mathrm{20\,nm}$ thick $\mathrm{Al_2O_3}$ etch-stop layer, a $\mathrm{279\,nm}$ thick $\mathrm{SiO_2}$ spacer and two $\mathrm{aSi/HfO_2}$ layer pairs on a semi-infinite crystalline-silicon substrate. Two layer pairs for the Bragg coating are the minimum number required to achieve a residual transmission below $\mathrm{10\,ppm}$, the non-reflected power of the full mirror is $\mathrm{1-R}\approx 6.6\cdot10^{-6}$. The estimated thermal noise amplitude spectral densities are shown in \textbf{Figure \ref{fig:noise_full}}. The modeling of thermal noise for the hybrid mirror is discussed in Appendix \ref{Appendix:thermalnoise}. The total etalon noise was calculated as the uncorrelated sum of all individual contributions according to Equation \ref{eq:uncorrelatedsumASD}. The non-reflected power of the mirror in dependence of the number of layer pairs as well as a schematic of the final structure are shown in \textbf{Figure \ref{fig:noise_DBR_sweep}}.

The total noise is dominated by the Brownian contribution of the $\mathrm{SiO_2}$ spacer across the entire frequency range, followed by the Brownian contribution of the Bragg coating. For the selected configuration, the estimated total thermal noise at a mechanical frequency of $\mathrm{100\,Hz}$ is approximately $\mathrm{1.97 \cdot 10^{-20}\,m/\sqrt{Hz}}$. Throughout the entire evaluated frequency range shown in Figure \ref{fig:noise_full}, the total noise remains approximately one order of magnitude below the projected ETpathfinder coating thermal noise. Since the total noise is calculated as an uncorrelated sum according to Equation \ref{eq:uncorrelatedsumASD} and therefore no correlations between thermo-elastic and thermo-refractive effects are considered, the calculated total noise amplitude represents a conservative estimate. 

        \begin{figure}[h!]
            \centering
            \begin{minipage}[c]{0.39\textwidth}
            \caption{
            Thermal-noise budget of the hybrid mirror with two $\mathrm{aSi/HfO_2}$ layer pairs, achieving a non-reflected power of $\mathrm{1-R}\approx 6.6\cdot10^{-6}$. The amplitude spectral densities of the Brownian, thermo-elastic and thermo-refractive noise contributions of the metasurface, Bragg coating, spacer and substrate are shown as functions of mechanical frequency. Solid, dashed and dotted lines represent Brownian, thermo-elastic and thermo-refractive noise, respectively. The dashed black line denotes the uncorrelated sum of all contributions, while the solid black line shows the projected ETpathfinder coating thermal noise for comparison. All calculations were performed at a temperature of 18 K and for a Gaussian beam radius of 2.2 mm.
            }
            \label{fig:noise_full}
            \end{minipage}
            \hfill
            \begin{minipage}[c]{0.6\textwidth}
                \centering
                \includegraphics[width=\linewidth]{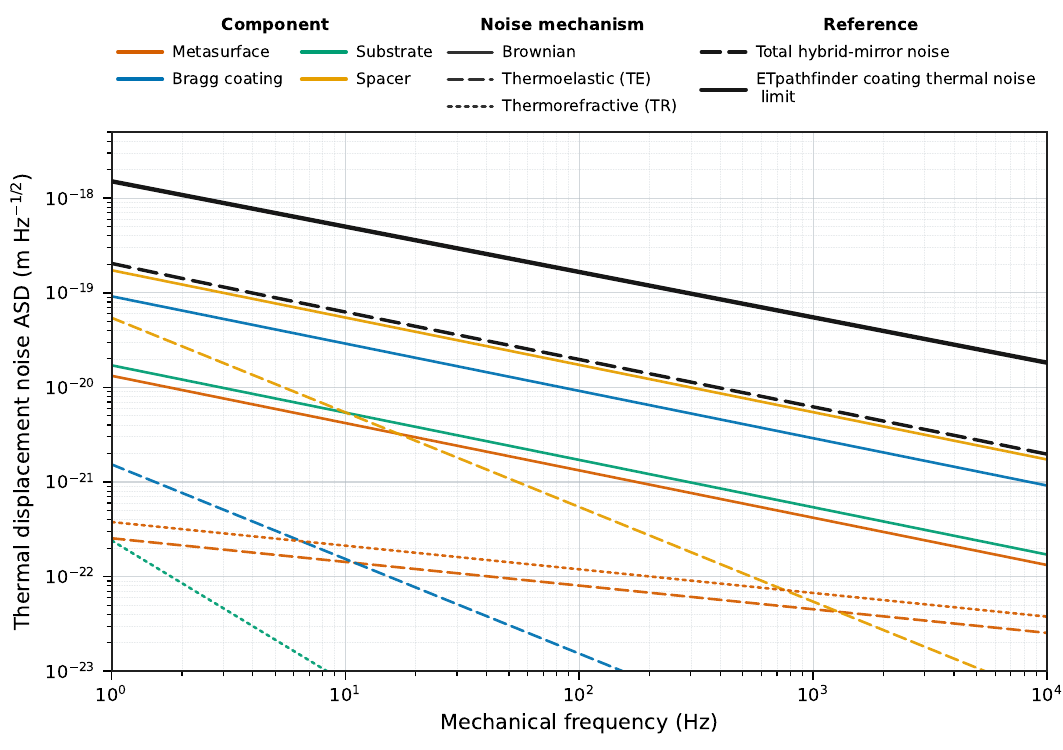}
            \end{minipage}
        \end{figure}

Figure \ref{fig:noise_DBR_sweep} a) shows the total thermal noise amplitude and non-reflected power of the mirror in dependence of the number of layer pairs used for the Bragg coating at a mechanical frequency of $\mathrm{100\,Hz}$. For low numbers of layer pairs for the Bragg coating the total noise is dominated by the Brownian contribution of the spacer. For 7 layer pairs, the Brownian contributions of the spacer and Bragg mirror are approximately equal, while for any layer pair count above 7 the Bragg coating provides the highest contribution to the total noise amplitude. Two layer pairs are the minimum required amount to achieve the desired residual transmission of below $\mathrm{10\,ppm}$. More than 6 Bragg layer pairs no longer have any effect on the non-reflected power, since it is limited by the absorption of the mirror. Figure \ref{fig:noise_DBR_sweep} b) shows a schematic of the final structure with all geometrical parameters as well as the normalized magnetic- and electric-field amplitudes throughout the entire structure. The fields are strongly localized inside and close to the metasurface ridges, while only a small fraction reaches into the Bragg mirror. This distribution illustrates the intended functionality of the hybrid mirror. The metasurface provides the dominant reflectance contribution, whereas the minimal Bragg coating compensates for the remaining transmission. The combination of both elements satisfies the transmission requirement, while reducing the estimated thermal-noise contribution associated with the multilayer coating. 

A comparison of the optical and noise performance of the hybrid mirror with a selection of different Bragg coatings is shown in \textbf{Table} \ref{tab:BraggVSHybrid}. The comparison indicates that the principal benefit of the hybrid architecture is not an exceptionally large noise reduction relative to every possible conventional coating, but an improved balance between optical loss and Brownian noise. In particular, the predicted Brownian noise reduction relative to the configuration made from aSi/$\mathrm{HfO_2}$ is approximately only $4.5\,\%$, while the modeled absorption is reduced by almost a factor of 24. A conventional $\mathrm{Ta_2O_5/SiO_2}$ coating achieves similar transmission and absorption values, but its Brownian noise amplitude is approximately 3.9 times higher. To provide a compact comparison of optical and thermal noise performance, we define an illustrative normalized combined figure of merit, $\mathrm{FOM_{comb}}$, incorporating residual transmission, absorption, and Brownian noise amplitude. The metric is normalized to the hybrid mirror, for which $\mathrm{FOM_{comb}} = 1$, with larger values indicating better combined performance:

    \begin{equation}
    \label{eq:FOM}
    FOM_\mathrm{{comb,j}} = \left[ \frac{T_\mathrm{j}}{T_\mathrm{hyb}} \frac{A_\mathrm{j}}{A_\mathrm{hyb}}  \frac{\sqrt{S_\mathrm{Brown, j}}}{\sqrt{S_\mathrm{Brown, hyb}}}  \right]^{-1/3} ,
    \end{equation}

where $j$ denotes the investigated configuration, $T$ is the transmission, $A$ is the absorption and $\sqrt{S_{\mathrm{Brown}}}$ is the Brownian noise amplitude according to Table \ref{tab:BraggVSHybrid}. This figure of merit is not intended as a universal performance criterion, since the relative importance of the three quantities depends on the specific application, but rather as an illustrative comparison under identical operating conditions.

    \begin{table}[h!]
    \centering
    \caption{Comparison of the optical and noise performance for the hybrid mirror and different Bragg mirror configurations. For each Bragg mirror, the minimum number of layer pairs required to satisfy $\mathrm{T_{res}\leq10^{-5}}$ was used. The optical properties of the Bragg coatings were calculated using the transfer-matrix method with the complex refractive indices listed in Table~\ref{table:matparam}, whereas those of the hybrid mirror were obtained from the full-wave FEM model. The optical properties of the Bragg coatings were additionally validated using the same FEM model. The structures are illuminated from air at normal incidence and deposited on a semi-infinite crystalline-silicon substrate. The Brownian noise calculations correspond to a temperature of $18\,\mathrm{K}$, a mechanical frequency of $100\,\mathrm{Hz}$, and a Gaussian beam radius of $2.2\,\mathrm{mm}$. The column ASD ratio gives the ratio between the Brownian noise ASD of each configuration and that of the hybrid mirror.}
    \label{tab:BraggVSHybrid}
    \begin{tabular}{lcccccc}
    \hline
    Layer materials &
    Layer pairs &
    Residual Transmission &
    Absorption &
    Brownian noise ASD &
    ASD ratio &
    $\mathrm{FOM_{comb}}$\\
    &
    &
    &
    &
    ($\mathrm{m/\sqrt{Hz}}$) &
    &
    \\
    \hline
    Hybrid mirror & 2 & 5.98 $\cdot 10^{-6}$ & 5.85 $\cdot 10^{-7}$ & 1.970 $\cdot 10^{-20}$ & 1 & 1 \\
    aSi/$\mathrm{HfO_2}$ & 10 & 5.91 $\cdot 10^{-6}$ & 1.39 $\cdot 10^{-5}$ & 2.062 $\cdot 10^{-20}$ & 1.05 & 0.344 \\
    aSi/$\mathrm{Ta_2O_5}$ & 10 & 5.45 $\cdot 10^{-6}$ & 7.67 $\cdot 10^{-6}$ & 3.527 $\cdot 10^{-20}$ & 1.79 & 0.360 \\
    aSi/$\mathrm{SiO_2}$ & 7 & 2.01 $\cdot 10^{-6}$ & 6.39 $\cdot 10^{-6}$ & 3.829 $\cdot 10^{-20}$ & 1.94 & 0.519\\
    $\mathrm{HfO_2}$/$\mathrm{SiO_2}$ & 18 & 6.03 $\cdot 10^{-6}$ & 3.07 $\cdot 10^{-5}$ & 6.697 $\cdot 10^{-20}$ & 3.40 & 0.177 \\
    $\mathrm{Ta_2O_5}$/$\mathrm{SiO_2}$ & 18 & 6.98 $\cdot 10^{-6}$ & 6.48 $\cdot 10^{-7}$ & 7.720 $\cdot 10^{-20}$ & 3.92 & 0.582 \\
    \hline
    \end{tabular}
    \end{table}

Under the material and optical-loss assumptions adopted here, the hybrid architecture provides the most favorable combined optical and thermal noise performance. However, the quantitative absorption advantage depends on the assumed aSi:H extinction coefficient and requires experimental confirmation, particularly with respect to fabrication-induced scattering.

\section{Conclusions}

We presented a fabrication-aware hybrid metasurface-Bragg mirror architecture that combines the low-noise potential of a resonant dielectric metasurface with the optical robustness of a reduced supporting Bragg coating. The metasurface provides the dominant reflectance contribution, while the spacer and Bragg coating are selected to establish the required phase relation and compensate the fabrication-limited residual transmission. Line-edge roughness was incorporated through an effective smoothing model, and simultaneous geometrical and material variations were evaluated using a truncated-Gaussian Monte Carlo analysis. Reoptimization with line-edge roughness included reduces the modeled $\mathrm{80\,\%}$-quantile non-reflected power from $6.22\cdot10^{-4}$ to $1.04\cdot10^{-4}$, corresponding to an improvement by a factor of approximately 6.

As a representative application example, the architecture was evaluated under the $1550\,\mathrm{nm}$ and $18\,\mathrm{K}$ operating conditions of ETpathfinder. For this case, an $\mathrm{aSi/HfO_2}$ Bragg coating was selected based on its estimated cryogenic Brownian noise. Two layer pairs are sufficient to reduce the non-reflected power of the complete mirror to approximately $6.6\,\mathrm{ppm}$. The estimated thermal displacement noise is $1.97\cdot10^{-20}\,\mathrm{m/\sqrt{Hz}}$ at $100\,\mathrm{Hz}$ and remains approximately one order of magnitude below the projected ETpathfinder coating thermal noise throughout the investigated frequency range. The total noise is dominated by the Brownian contribution of the SiO$_2$ spacer, followed by the noise contribution of the reduced Bragg mirror. The noise budget identifies spacer thickness and cryogenic mechanical loss as key design parameters alongside the optical performance of the metasurface and supporting Bragg coating.

These results show that the residual transmittance of fabrication-limited metasurfaces can be suppressed by a minimal supporting Bragg coating while retaining a low thermal-noise amplitude. The presented framework directly connects fabrication tolerances, optical performance, and thermal-noise reduction and can be adapted to other fabrication processes and tolerance ranges. The concept establishes a design route toward high-reflectivity, low-noise optics for next-generation gravitational-wave detectors, including the Einstein Telescope, and other precision optical systems such as optical clocks and ultra-stable cavities.

\medskip

% Acknowledgements
\medskip
\textbf{Acknowledgments} \par %delete if not applicable))
     Part of this work was supported by the \textit{German Federal Ministry of Research, Technology and Space} (BMFTR), chapter 4, title 89450, fiscal year 2024, project: ''Verbundprojekt 05A2023 - 3G-GWD: Gravitationswellenteleskope der dritten Generation. Teilprojekt 12.'', funding ID number 05A24SJ1. The authors acknowledge partial support by the Deutsche Forschungsgemeinschaft (DFG, German Research Foundation) under Germany’s Excellence Strategy – EXC 2123/2 QuantumFrontiers – 390837967 and support by the German Federal Ministry of Research, Technology and Space (BMFTR) under grant number 05A2023. The authors thank Liam Shelling Neto for providing RCWA code and for valuable discussions on numerical convergence.

\section*{Conflict of Interest}
The authors declare no conflict of interest.

\section*{Data Availability Statement}
The data that support the findings of this study are available from the corresponding author upon reasonable request.

\section*{Generative AI Use Statement}
During the preparation of this work, the authors used OpenAI ChatGPT for language editing and assistance with the preparation of plotting code. All scientific content, calculations, data, interpretations, and final wording were reviewed and verified by the authors, who take full responsibility for the content of the manuscript.

\appendix
\newpage
\section{Material Parameters}
\label{Appendix:material_parameters}

\begin{table*}[h!]
    \centering
    \caption{Material parameters used for the calculations at $\mathrm{T=18\,K}$: Young's modulus $Y$, Poisson's ratio $\nu$,
    density $\rho$, thermal conductivity $\kappa$, specific heat $C$,
    thermal expansion coefficient $\alpha$, thermo-optic coefficient
    $\beta$, refractive index $n$, extinction coefficient $k$, and
    mechanical loss factor $\phi$.}
    \label{table:matparam}

    \fontsize{6.5pt}{7.5pt}\selectfont
    \setlength{\tabcolsep}{2pt}
    \renewcommand{\arraystretch}{1.15}

    \begin{tabularx}{\textwidth}{
        >{\raggedright\arraybackslash}p{1.65cm}
        *{6}{C}
    }
        \toprule
        Parameter
        & c-Silicon
        & a-Silicon
        & $\mathrm{SiO_2}$
        & $\mathrm{HfO_2}$
        & $\mathrm{Ta_2O_5}$
        & a-$\mathrm{Al_2O_3}$ \\
        \midrule

        $Y$ ($\mathrm{GPa}$)
        & 130 \cite{nawrodt2009thermal}
        & 140 \cite{follstaedt2004mechanical}
        & 72 \cite{heinert2013calculation}
        & 165 \cite{berdova2016hardness}
        & 140 \cite{alcala2002mechanical}
        & 122 \cite{alcala2002mechanical} \\

        $\nu$
        & 0.22 \cite{mpdbJAHM}
        & 0.22 \cite{freund2004thin}
        & 0.159 \cite{simon1994cryogenic}
        & 0.3 \cite{berdova2016hardness}
        & 0.21 \cite{franc2009mirrorthermalnoiselaser}
        & 0.2 \cite{franc2009mirrorthermalnoiselaser} \\

        $\rho$ ($\mathrm{kg/m^{3}}$)
        & 2331 \cite{mpdbJAHM}
        & 2331 \cite{mpdbJAHM}
        & 2200 \cite{braginsky2003thermodynamical}
        & 8000 \cite{franc2009mirrorthermalnoiselaser}
        & 6850 \cite{gorodetsky2008thermal}
        & 3700 \cite{franc2009mirrorthermalnoiselaser} \\

        $\kappa$ ($\mathrm{W/(m\,K})$)
        & 2330 \cite{mpdbJAHM}
        & 0.2 \cite{zink2006thermal}
        & 0.13 \cite{simon1994cryogenic}
        & 0.2 \cite{zeng2025thermal}
        & 0.6 \cite{welsch1999absolute}
        & 0.2\textsuperscript{*} \cite{harper2024vibrational} \\

        $C$ ($\mathrm{J/(kg\,K)}$)
        & 0.276 \cite{mpdbJAHM}
        & 3.5 \cite{zink2006thermal}
        & 7 \cite{simon1994cryogenic}
        & 0.378 \cite{zhou2011heat}
        & 3.17 \cite{touloukian1970specific}
        & 10\textsuperscript{*}
          \cite{etingergeller2019thickness,zeller1971thermal} \\

        $\alpha$ ($\mathrm{1/K}$)
        & $4.85 \cdot 10^{-10}$ \cite{mpdbJAHM}
        & $1 \cdot 10^{-7}$ \cite{fabian1997thermal}
        & $-2.5 \cdot 10^{-7}$ \cite{franc2009mirrorthermalnoiselaser}
        & $3.8 \cdot 10^{-7}$\textsuperscript{*}
          \cite{touloukian1977thermalexpansion}
        & $3.6 \cdot 10^{-7}$\textsuperscript{*}
          \cite{gorodetsky2008thermal}
        & $6 \cdot 10^{-7}$ \cite{touloukian1977thermalexpansion} \\

        $\beta$ ($\mathrm{1/K}$)
        & $5.8 \cdot 10^{-6}$ \cite{freyTempDependentSiliconGermanium}
        & $1 \cdot 10^{-6}$ \cite{komma2012thermo}
        & $1.01 \cdot 10^{-6}$ \cite{franc2009mirrorthermalnoiselaser}
        & $2.44 \cdot 10^{-5}$ \cite{jaramillo2025hfo2}
        & $2.3 \cdot 10^{-6}$ \cite{tien2000simultaneous}
        & $1.3 \cdot 10^{-5}$
          \cite{tomaru2002thermal,saleem2013thermal} \\

        $n$ at $\mathrm{1550\,nm}$
        & 3.45 \cite{freyTempDependentSiliconGermanium}
        & 3.774 (measured); 3.6211 (a-Si:H)
          \cite{franta2013advanced}
        & 1.4635 \cite{franta2016optical}
        & 2.052 \cite{franta2015universal}
        & 2.0436 \cite{franta2015dispersion}
        & 1.619 (measured) \\

        $k$ at $\mathrm{1550\,nm}$
        & $3.3 \cdot 10^{-9}$ \cite{steinlechner2013optical}
        & $1.22 \cdot 10^{-5}$(a-Si)
          \cite{birney2018amorphous};
          $1.3 \cdot 10^{-7}$(a-Si:H)
          \cite{franta2013advanced, Takei:14}
        & $1.0973 \cdot 10^{-7}$ \cite{franta2016optical}
        & $1 \cdot 10^{-5}$ \cite{Nunez:23}
        & $1 \cdot 10^{-7}$ \cite{Belt:17}
        & $3 \cdot 10^{-7}$ \cite{Al2O3_absorption} \\

        $\phi$ (rad)
        & $1 \cdot 10^{-9}$ \cite{mpdbJAHM}
        & $1 \cdot 10^{-5}$ \cite{murray2015ion}
        & $5 \cdot 10^{-4}$ \cite{franc2009mirrorthermalnoiselaser}
        & $2.2 \cdot 10^{-4}$ \cite{chalkley2008mechanical}
        & $6 \cdot 10^{-4}$ \cite{martin2010effect}
        & $5 \cdot 10^{-4}$ \cite{fazio2022comprehensive} \\

        \bottomrule
    \end{tabularx}

    \vspace{2pt}
    \begin{minipage}{\textwidth}
        \scriptsize
        \textsuperscript{*}Values were estimated or extrapolated from
        the cited references.
    \end{minipage}
\end{table*}

\section{Bragg-Mirror selection based on Brownian thermal noise}
\label{Appendix:Bragg}

    As a figure of merit for evaluating different Bragg mirror configurations, their total Brownian thermal noise is considered. Brownian noise generally represents one of the dominant thermal noise contributions in high-reflectivity dielectric mirror coatings and therefore provides a suitable metric for an initial comparison of mirror performance. Thermo-optical effects, including thermo-elastic and thermo-refractive noise, are not included at this stage and are instead considered later within the complete etalon noise analysis of the selected design.
    The Brownian displacement noise is calculated using the thin-coating, semi-infinite-substrate model derived by Harry et al. \cite{harry2002}. Treating the mechanical loss of each coating material as isotropic, assuming identical loss angles for strains parallel and perpendicular to the coating surface, the one-sided displacement-noise power spectral density can be written as:

    \begin{equation}
    S_\mathrm{x}(f) = \frac{2k_{\mathrm{B}}T}{\pi^{3/2}f} \frac{1-\nu_{\mathrm{s}}^{2}}{r_0Y_{\mathrm{s}}} \phi_{\mathrm{s}} +
    \frac{2k_{\mathrm{B}}T}{\pi^2 f r_0^2} \sum_{i=1}^{N_{\mathrm{lay}}} d_\mathrm{i}\phi_\mathrm{i} \left[ \frac{Y_\mathrm{i}}{Y_{\mathrm s}^{2}}
    \frac{(1+\nu_{\mathrm{s}})^2(1-2\nu_{\mathrm{s}})^2} {1-\nu_\mathrm{i}^2} + \frac{1}{Y_\mathrm{i}} \frac{(1+\nu_\mathrm{i})^2(1-2\nu_\mathrm{i})} {1-\nu_\mathrm{i}^2} \right].
    \label{eq:BraggBrownian}
    \end{equation}

    Here, $S_\mathrm{x}(f)$ is the one-sided displacement-noise power spectral density at the mechanical frequency $f$, $k_{\mathrm B}$ is the Boltzmann constant, $T$ is the temperature, and $r_\mathrm{0}$ is the radius of the Gaussian beam. The quantities $Y_{\mathrm s}$, $\nu_{\mathrm s}$, and $\phi_{\mathrm s}$ denote the Young’s modulus, Poisson’s ratio, and mechanical loss angle of the substrate, respectively. The index $i$ refers to an individual coating layer, while $N_{\mathrm{lay}}$ is the total number of coating layers. The quantities $d_\mathrm{i}$, $Y_\mathrm{i}$, $\nu_\mathrm{i}$, and $\phi_\mathrm{i}$ are the thickness, Young’s modulus, Poisson’s ratio, and mechanical loss angle of layer $i$, respectively. The first term in Equation \ref{eq:BraggBrownian} describes substrate Brownian noise, whereas the sum by layer gives the coating contribution. The corresponding amplitude spectral density is obtained as $\sqrt{S_x(f)}$.
    Equation \ref{eq:BraggBrownian} is the isotropic-loss specialization of Eq. (22) of Harry et al. \cite{harry2002}, applied separately to each layer of the coating stack. The calculation assumes perfectly bonded, mechanically isotropic layers that are thin compared with the beam radius and deposited on a homogeneous semi-infinite substrate. The virtual pressure used to evaluate the displacement response has the same Gaussian profile as the laser beam for a specific application. Each layer is assumed to be of quarter-wave optical thickness:

    \begin{equation}
    \label{eq:quarterwave}
    d_\mathrm{i}=\frac{\lambda_0}{4n_\mathrm{i}},
    \end{equation}

    where $\lambda_0=1550\,\mathrm{nm}$ is the vacuum wavelength and $n_\mathrm{i}$ is the real part of the refractive index of layer $i$.

    \noindent Only Bragg mirror designs with a residual transmission of $\mathrm{T_{res}\leq10^{-5}}$ are considered further. \textbf{Table} \ref{tab:BraggComparison} summarizes the investigated configurations together with their total layer count and calculated coating Brownian noise amplitude spectral density. The substrate material for all configurations was crystalline silicon. All material parameters used for the calculations can be found in Table \ref{table:matparam}. 

    \begin{table}[ht]
    \centering
    \caption{Comparison of Bragg-mirror configurations using the minimum number of quarter-wave layer pairs required to satisfy $\mathrm{T_{res}\leq10^{-5}}$. The optical properties were calculated using the transfer-matrix method with the complex refractive indices listed in Table~\ref{table:matparam}. The structures are illuminated from air at normal incidence and deposited on a semi-infinite crystalline-silicon substrate. The Brownian noise calculations correspond to a temperature of $18\,\mathrm{K}$, a mechanical frequency of $100\,\mathrm{Hz}$, and a Gaussian beam radius of $2.2\,\mathrm{mm}$.}
    \label{tab:BraggComparison}
    \begin{tabular}{lcccc}
    \hline
    Layer materials &
    Layer pairs &
    Residual Transmission &
    Absorption &
    Brownian noise ASD \\
    &
    &
    &
    &
    ($\mathrm{m/\sqrt{Hz}}$) \\
    \hline
    aSi/$\mathrm{HfO_2}$ & 10 & 5.91 $\cdot 10^{-6}$ & 1.39 $\cdot 10^{-5}$ & 2.062 $\cdot 10^{-20}$ \\
    aSi/$\mathrm{Ta_2O_5}$ & 10 & 5.45 $\cdot 10^{-6}$ & 7.67 $\cdot 10^{-6}$ & 3.527 $\cdot 10^{-20}$ \\
    aSi/$\mathrm{SiO_2}$ & 7 & 2.01 $\cdot 10^{-6}$ & 6.39 $\cdot 10^{-6}$ & 3.829 $\cdot 10^{-20}$ \\
    $\mathrm{HfO_2}$/$\mathrm{SiO_2}$ & 18 & 6.03 $\cdot 10^{-6}$ & 3.07 $\cdot 10^{-5}$ & 6.697 $\cdot 10^{-20}$ \\
    $\mathrm{Ta_2O_5}$/$\mathrm{SiO_2}$ & 18 & 6.98 $\cdot 10^{-6}$ & 6.48 $\cdot 10^{-7}$ & 7.720 $\cdot 10^{-20}$ \\
    \hline
    \end{tabular}
    \end{table}

    \noindent According to the results shown in Table \ref{tab:BraggComparison}, the configuration made from amorphous silicon and hafnia is the most promising and will be used for the final design. Even though the configuration made from amorphous silicon and silicon dioxide has significantly less total layers, the high mechanical loss factor of silicon dioxide at low temperatures results in a higher noise amplitude. Mirrors made from hafnia and silicon dioxide as well as tantala and silicon dioxide have the highest noise amplitudes in this comparison, as they are made up of a large amount of total layers due to the low refractive index contrast between the two materials. In addition to this, the high mechanical loss factor of the materials at cryogenic temperatures also significantly increases the overall noise of these configurations.

    As an independent validation, the ten-pair aSi/$\mathrm{HfO_2}$ mirror was also evaluated using an axisymmetric finite-element model with a Gaussian virtual pressure and a substrate radius and depth of $50\cdot r_\mathrm{0}$, with $r_\mathrm{0}=2.2\,\mathrm{mm}$ being the radius of the Gaussian virtual pressure. The simulated total Brownian noise amplitude was $2.056\cdot10^{-20}\,\mathrm{m/\sqrt{Hz}}$, which differs by approximately $0.3\%$ from the analytical value of $2.062\cdot10^{-20}\,\mathrm{m/\sqrt{Hz}}$. This agreement confirms the implementation of Equation \ref{eq:BraggBrownian} and the numerical convergence of the finite-element half-space approximation.
   
\section{Near-field-constrained spacer selection}
\label{Appendix:spacer}
        Since the metamirror and Bragg-mirror form an etalon, the spacer thickness has to satisfy the antiresonance condition at the design wavelength \cite{Vaughan1989FabryPerot}:

        \begin{equation} 
        d_{\mathrm{spacer}} = \frac{1}{n_{\mathrm{spacer}}} \left[ \frac{(2m+1)\lambda_0}{4} - \frac{\lambda_0}{4\pi} \left( \Phi_{\mathrm{upper}} + \Phi_{\mathrm{DBR}} \right)\right], 
        \label{eq:spacer_thickness_antiresonance} 
        \end{equation}        

        \noindent where $\lambda_0$ is the vacuum wavelength, $n_\mathrm{spacer}$ is the refractive index of the spacer material, $\Phi_\mathrm{upper}$ is the phase added from a reflection of the top layer made from the metasurface and etch-stop layer, $\Phi_\mathrm{DBR}$ is the phase added from a reflection of the Bragg mirror and $m$ is an integer number. The phases were evaluated using FEM simulations and are $\mathrm{\Phi_{upper} = 0.9449\,\pi\, rad}$ and $\mathrm{\Phi_{DBR} = -0.9994\,\pi\, rad}$, respectively. The lowest-order solution for $\mathrm{m=0}$ represents the smallest possible geometrical thickness. Care has to be taken when choosing the spacer dimension, because for thicknesses too small, the evanescent field behind the metasurface will couple to the first layer of the Bragg-mirror and therefore significantly increase the overall transmittance. We therefore numerically evaluate all possible spacer lengths given by Equation \ref{eq:spacer_thickness_antiresonance} for successively increasing orders. No significant decrease in the total transmittance of the mirror could be observed for increasing orders. Therefore, the smallest spacer thickness for $\mathrm{m=0}$ was chosen, resulting in a spacer thickness of $\mathrm{d_{spacer} \approx 279\,nm}$.

\section{Thermal noise modeling for hybrid mirror}
\label{Appendix:thermalnoise}

        This section introduces the thermal noise model used to estimate the thermal displacement noise of the hybrid mirror. Thermal noise in optical coatings can mainly be divided into two categories: Brownian thermal noise and thermo-optical noise. Brownian thermal noise originates from mechanical dissipation within the coating materials \cite{harry2002, Hong2013}. At a finite temperature, thermal energy excites mechanical degrees of freedom \cite{Levin1998}, leading to a stochastic displacement of the coating layer surfaces, as well as dimensional changes of the layers.
        Thermo-optical noise, in contrast, arises from thermal dissipation within the coating material \cite{Evans2008}. These processes cause temperature fluctuations that induce variations in both the coating dimensions and refractive index. Thermo-optical noise can be further separated into thermo-elastic noise, associated with the thermally driven dimensional changes, and thermo-refractive noise, associated with temperature-dependent refractive index fluctuations \cite{Braginsky2003}. Together, these effects lead to spatial and temporal variations of the optical path length and therefore to phase fluctuations in the reflected and transmitted optical fields. 
        Individual thermal noise contributions are evaluated separately for the relevant system components and combined through an uncorrelated sum:
    
        \begin{equation}
        \label{eq:uncorrelatedsumASD}
        \sqrt{S_{\mathrm{tot}}} = \sqrt{\sum_{i} S_{i}},
        \end{equation}    
    
        assuming statistical independence between noise sources. This provides a conservative approximation for the total thermal noise budget. Generally, correlations between certain mechanisms, particularly thermo-elastic and thermo-refractive noise, can lead to partial compensation effects that reduce the overall thermo-optical noise contribution \cite{Evans2008}. \\
        \noindent The calculation of all thermal noise contributions for the complete etalon system closely follows the scheme described in reference \cite{dickmann2018b} with some exceptions. The Brownian, thermo-elastic, and thermo-refractive noise contributions of the metasurface, spacer, Bragg-mirror and substrate are calculated separately and then summed according to Equation \ref{eq:uncorrelatedsumASD}.

        \subsection{Brownian thermal noise}
    
    The calculation of Brownian thermal noise for the metasurface is based on a formulation of the fluctuation-dissipation-theorem \cite{Callen195134} by Levin \cite{Levin1998} that was adapted for the application of functional optical surfaces \cite{kroker2017a, dickmann2018c}. It enables the calculation of Brownian noise for metasurfaces of arbitrary shape \cite{Gaedtke2026BrownianNoise}. \\
    \noindent The approach uses a virtual pressure, which is oscillating at a mechanical frequency of interest and is applied to the nanostructured surface. Elastic energy is introduced into the material as a result, which leads to a dissipated power:

    \begin{equation}
    \label{eq:Pdiss}
    W_{\mathrm{diss}}(\omega) = \omega \int \epsilon(\vec{r})\phi(\vec{r})\,dV,
    \end{equation}

    \noindent where $\omega$ is the mechanical angular frequency of the oscillation, $\vec{r}$ is a location on the surface, $\epsilon(\vec{r})$ is the introduced elastic energy density and $\phi(\vec{r})$ is the mechanical loss factor of the coating. The dissipated power can then be used to calculate the power spectral density of Brownian thermal displacement noise \cite{Levin1998}:

    \begin{equation}
    \label{eq:S_PSD}
    S(\omega, T) = \frac{8 k_\mathrm{B} T}{\omega^2}\frac{W_{\mathrm{diss}}}{F_0^2},
    \end{equation}

    \noindent with Boltzmann constant $k_\mathrm{B}$, temperature $T$ and $F_0$ the total force applied to the surface. The shape of this applied force is solely determined by the distribution of electromagnetic energy within the structure \cite{kroker2017a, TUGOLUKOV20182181}. The electric and magnetic fields inside the structure lead to a ponderomotive pressure of the form:

    \begin{equation}
    \label{eq:ponderomotivepressure}
    p_{\perp}(\vec{r}) = \sum_{\mathrm{ij}}\Delta \sigma_{\mathrm{ij}}(\vec{r})n_\mathrm{i}         n_\mathrm{j},
    \end{equation}
    
    \noindent where $\Delta \sigma_{\mathrm{ij}}$ is the difference between the corresponding Maxwell-stress-tensor components evaluated on the two sides of an interface, $i$ and $j$ are the spatial coordinate basis and $n$ is the normal vector of a given surface. The Maxwell stress tensor is given by:

    \begin{equation}
    \label{eq:Maxwellstresstensor}
    \sigma_{\mathrm{ij}} = \epsilon_\mathrm{0} \epsilon_\mathrm{r} E_\mathrm{i} E_\mathrm{j} + \frac{1}{\mu_\mathrm{0} \mu_\mathrm{r}}B_\mathrm{i} B_\mathrm{j} - \frac{1}{2}\left(\epsilon_\mathrm{0} \epsilon_\mathrm{r} E^2 + \frac{1}{\mu_\mathrm{0} \mu_\mathrm{r}} B^2 \right) \delta_{\mathrm{ij}},
    \end{equation}
    
    \noindent with the dielectric and magnetic field constants $\epsilon_\mathrm{0}$ and $\mu_\mathrm{0}$, the vacuum amplitude of the electric field $E_\mathrm{i}$, the amplitude of the magnetic field $B_\mathrm{i}$ and the Kronecker symbol $\delta_\mathrm{{ij}}$. \\
    For the calculation of the Brownian noise contribution of the spacer, Bragg coating, and substrate we adopt the separation of spatial scales recently introduced by Pandey et al. \cite{Pandey2026} for hybrid metasurface-multilayer mirrors. Since the metasurface period is much smaller than the Gaussian beam radius, the Maxwell pressure at each material interface can be separated into a period-averaged component and a zero-mean component varying over the metasurface period. These two pressure components produce elastic responses on different spatial scales and therefore require separate mechanical models.

    The period-averaged contribution describes the deformation extending over the full beam radius. It is evaluated using an axisymmetric finite-element model of the complete layer stack, in which the metasurface is represented by a homogeneous effective layer. A Gaussian virtual pressure with a total force of $1\,\mathrm{N}$ and a $1/e^2$ intensity radius of $r_\mathrm{0} = 2.2\,\mathrm{mm}$ was applied at the effective reflection plane located at the upper surface of the etch-stop layer. The Gaussian pressure represents the slowly varying beam envelope and does not imply that the local electromagnetic field at the underlying interfaces has a Gaussian distribution. The substrate radius and depth were both set to $50\cdot \mathrm{r_{0}}$, for which the calculated elastic energies were found to be converged. This provides a finite-element approximation of the infinite half-space considered by Harry et al. \cite{harry2002} and Hong et al. \cite{Hong2013}. The implementation was independently validated using the conventional aSi/HfO$_\mathrm{2}$ Bragg mirror with ten layer pairs as discussed in Appendix \ref{Appendix:Bragg}.

    The zero-mean, period-scale contribution was calculated using a two-dimensional periodic unit cell of the complete hybrid mirror. The local Maxwell pressures were obtained from the electromagnetic field distribution of the full structure. For each horizontal interface, the period-averaged Maxwell pressure was subtracted and the remaining zero-mean pressure was applied. This treatment retains the local field enhancement, the spatial variation and sign of the interface forces, and the resulting elastic deformation of the metasurface. The elastic energies per unit invariant length obtained from the unit-cell simulation were converted to the response of the full circular Gaussian beam using the integration described by reference \cite{Pandey2026}. The normalization was determined from the net optical force corresponding to the incident optical power and the simulated reflectance and transmittance of the hybrid mirror.

    Since the elastic response is linear, the complete deformation is the sum of the smooth and periodic responses. The corresponding elastic energy additionally contains a signed cross term \cite{Pandey2026}, which was calculated from the domain-wise overlap between the smooth strain tensor and the periodic stress tensor. For the specific design investigated in section \ref{sec:pathfinder}, the cross term was found to be $-1.15\cdot10^{-42}\,\mathrm{m^2/Hz}$ at $\mathrm{100\,Hz}$. It therefore accounts only for approximately $-0.3\,\%$ of the total power spectral density associated with the Brownian noise. The period-averaged, periodic and cross contributions were each weighted by the mechanical loss angle of the respective materials and combined to a displacement power spectral density. This procedure was applied separately to the metasurface, the etch-stop layer, the spacer layer, the Bragg coating and the substrate layer, allowing their individual contributions to the complete Brownian noise to be determined.

        \subsection{Thermo-elastic noise}
        The calculation of thermo-elastic noise is based on the fluctuation-dissipation-theorem as well. The dissipation mechanism of this type of noise is heat flow within the materials caused by local volume fluctuations \cite{dickmann2018b}. In this case the dissipated power in Equation \ref{eq:S_PSD} is given by \cite{Liu2000thermoelasticnoise, dickmann2018b}:

            \begin{equation}
            \label{eq:W_diss_TE}
            W_{\mathrm{diss}} = 2\pi \kappa T \left(\frac{Y \alpha}{(1-2\nu)C \rho} \right)^2\int_h\int_0^R [\nabla\theta]^2 rdr,
            \end{equation}

        with the thermal conductivity $\kappa$, the specific heat capacity $C$, the Young's modulus $Y$, the Poisson's ratio $\nu$, the thermal expansion coefficient $\alpha$, the density $\rho$ and the trace of the strain tensor $\theta$. Equation \ref{eq:W_diss_TE} is used to calculate the noise contributions of the spacer, Bragg mirror and substrate, respectively.
        \noindent The contribution of the metasurface has to be determined by rigorous coupled-wave analysis (RCWA) \cite{moharam1981a} as described in \cite{heinert2013calculation, dickmann2018b}. A general temperature change of $\mathrm{\Delta T}$ will cause a relative length change of the respective material of $\mathrm{\delta = \alpha \Delta T}$. For small temperature changes, the resulting change of the reflected phase has a linear relationship \cite{dickmann2018b, heinert2013calculation}:

            \begin{equation}
            \label{eq:K_TE}
            \delta\varphi = K_{\mathrm{TE}}\delta.
            \end{equation}   

        \noindent Here, $K_\mathrm{TE}$ is a proportional factor determined from the RCWA simulation. Contributions that need to be considered are the geometrical change of the metasurface dimensions, the effective movement of the surface towards the incident light and the change in metasurface period \cite{dickmann2018b}. The metasurface contribution to TE noise is then given by \cite{heinert2013calculation}:

            \begin{equation}
            \label{eq:S_TE_meta}
            S_{\mathrm{TE}}^{meta} = \left(\frac{\lambda}{4 \pi} K_{\mathrm{TE}} \alpha\right)^2 S_{T},
            \end{equation}   

        where $S_\mathrm{T}$ is the noise power of temperature fluctuations \cite{braginsky2003thermodynamical}:

            \begin{equation}
            \label{eq:S_T}
            S_{T} = \frac{k_B T^2}{\pi^{3/2}r_0^2 \sqrt{\rho C \kappa f}}.
            \end{equation}           

        \subsection{Thermo-refractive noise}

        Thermo-refractive noise originates from local fluctuations of the refractive index with temperature \cite{gurkovsky2011a}. For this noise type, only the metasurface, spacer and Bragg-mirror need to be considered. The power spectral density of the thermo-refractive noise of the spacer is given by \cite{gurkovsky2011a}:

            \begin{equation}
            \label{eq:S_spacer_TR}
            S_{\mathrm{TR}}^{spacer} = e_2^2\frac{k_B T^2 \beta^2 \kappa d}{\pi^2 \rho^2 C^2 r_0^4f^2} \left( 1 + \frac{1}{1+(4\pi / \lambda \sqrt{\kappa / 2\pi C \rho f })^4} \right),
            \end{equation}          

        \noindent where $\beta$ is the thermo-optic coefficient, $T$ is the temperature, $\kappa$ is the thermal conductivity, $d$ is the spacer thickness, $\rho$ is the density, $C$ is the specific heat, ${r_0}$ is the beam radius, $f$ is the mechanical frequency and $\lambda$ is the wavelength. The second term originates from the standing wave inside of the etalon \cite{dickmann2018b}. The coefficient $\mathrm{e_2}$ was introduced in reference \cite{dickmann2018b} and scales the noise contribution with the reflectances of the front and back mirror of the etalon. It is given by:

            \begin{equation}
            \label{eq:e2}
            e_2=\frac{n_\mathrm{s}\sqrt{R_2}(1-R_1)}{\left(1+\sqrt{R_1R_2}\right)^2},
            \end{equation}
        
        \noindent where $R_1$ is the reflectance of the metasurface, $R_2$ is the reflectance of the Bragg-mirror, and $n_\mathrm{s}$ is the refractive index of the spacer. For the Bragg-mirror, the power spectral density of thermo-refractive noise is given by \cite{gurkovsky2011a}:

            \begin{equation}
            \label{eq:S_coating_TR}
            S_{\mathrm{TR}}^{coating} = e_2^2\frac{k_B T^2 \beta_{\mathrm{eff}} \lambda_0^2}{\pi^{3/2} r_0^2 \sqrt{\kappa \rho C f}},
            \end{equation} 

            with \cite{gurkovsky2011a}:

            \begin{equation}
            \label{eq:beta_eff}
            \beta_{\mathrm{eff}} = \frac{1}{4}\frac{\beta_1 n_2^2 + \beta_2 n_1^2}{n_1^2-n_2^2},
            \end{equation}             

            \noindent the effective thermo-optic coefficient of the coating. Here, the indices 1 and 2 correspond to the two materials used for the coating stack. The contribution of the metasurface to thermo-refractive noise is determined similarly to the contribution of thermo-elastic noise. Generally, a change in the refractive index of the metasurface material changes the phase of the reflected light. For small changes, the phase change is proportional to the change of refractive index \cite{dickmann2018b}:

            \begin{equation}
            \label{eq:K_TR}
            \delta\varphi = K_{\mathrm{TR}} \Delta n,
            \end{equation}             

            where $K_\mathrm{TR}$ is again a proportional constant. This constant is again numerically evaluated using RCWA simulations. The power spectral density of noise is then given by \cite{dickmann2018b, heinert2013calculation}:

            \begin{equation}
            \label{eq:S_TR}
            S_{\mathrm{TR}}^{meta} = \left( \frac{\lambda}{4 \pi} K_{\mathrm{TR} \beta} \right)^2 S_{\mathrm{T}},
            \end{equation}    

            where $S_\mathrm{T}$ again represents the noise power of temperature fluctuations.

% References
\newpage

% Use the following code if you wish to generate your bibliography with BibTeX;
% replace the string "MSP-template" below with the name(s) of
% the BibTeX data base(s) you want to use.
% The resulting bibliography-output (the content of the .bbl file)
% must be pasted back into this file before submission.
% Please also include your BibTeX data base file(s) in your submission
% so that we can re-run BibTeX if necessary.
%
\bibliographystyle{MSP}
\bibliography{20250103_MetaMirrors.bib}

\clearpage
\markboth{TABLE OF CONTENTS}{TABLE OF CONTENTS}
\section*{Table of Contents}

\noindent
A fabrication-aware hybrid mirror combines a resonant metasurface with a
reduced Bragg coating to achieve high reflectance using less coating material.
Accounting for line-edge roughness and simultaneous manufacturing variations
yields a robust design. In a cryogenic precision interferometry case study,
two Bragg layer pairs reduce the non-reflected power to
$6.6\,\mathrm{ppm}$ while retaining low thermal noise.

\vspace{1em}

\begin{center}
    \includegraphics[
        width=55mm,
        height=50mm,
        keepaspectratio
    ]{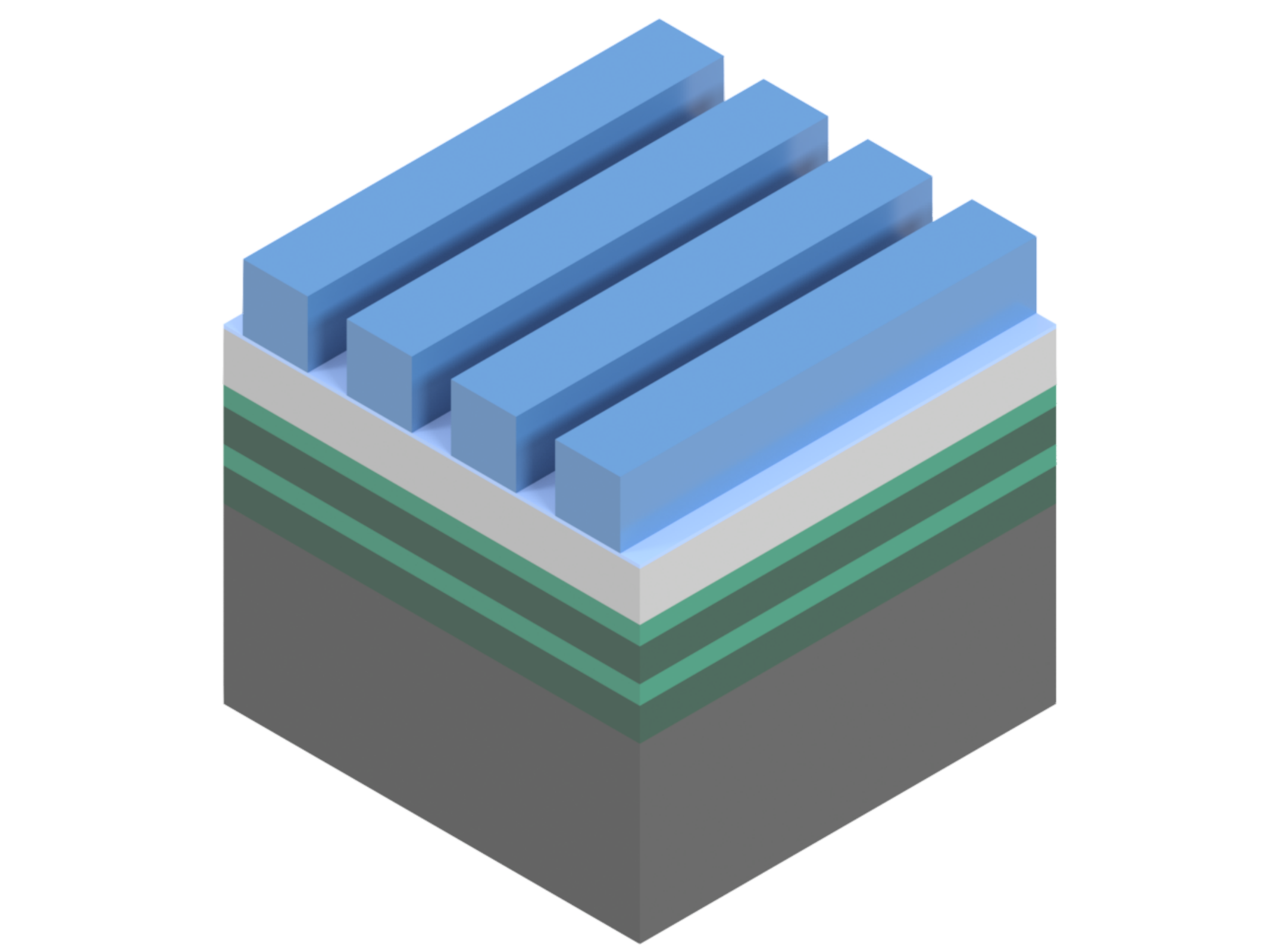}
\end{center}

\end{document}